\documentstyle[preprint,aps]{revtex}	
\begin{document}
%
%
\begin{titlepage}

\begin{flushright}
{
FSU--HEP--940307\\
UCD--94--5\\
hep-ph/9403248\\
March 1994\\
}
\end{flushright}

\vspace{5mm}

\begin{center}

{\Large\bf Amplitude Zeros in $W^\pm Z$ Production}

\vspace{7mm}

{\bf U.~Baur}\\[1mm]
{\it Department of Physics, Florida State University,
     Tallahassee, FL 32306, USA}\\[4mm]

{\bf T.~Han and J.~Ohnemus}\\[1mm]
{\it Department of Physics, University of California,
     Davis, California 95616, USA}

\end{center}

\vspace{7mm}

\begin{abstract}
We demonstrate that the Standard Model amplitude for
$f_1 \bar f_2 \rightarrow W^\pm Z $ at the Born-level exhibits an
approximate zero located at  $\cos\theta
=  (g^{f_1}_{-} + g^{f_2}_{-}) /  (g^{f_1}_{-} - g^{f_2}_{-})$  at high
energies, where the $g^{f_i}_{-}$ ($i=1,2$) are the left-handed couplings of
the $Z$-boson to fermions and $\theta$ is the center of mass scattering
angle of the $W$-boson. The approximate zero is the combined result of
an exact zero in the dominant helicity amplitudes ${\cal M}(\pm,\mp)$ and
strong gauge cancelations in the remaining amplitudes.
For non-standard $WWZ$ couplings these cancelations no longer occur and the
approximate amplitude zero is eliminated.
\end{abstract}

\end{titlepage}

%
%
\narrowtext

Although the electroweak Standard Model (SM) based on an
${\rm SU_L(2)}\bigotimes {\rm U_Y(1)}$
gauge theory  has been very successful
in describing contemporary high  energy physics experiments, the three
vector-boson couplings predicted by this non-Abelian gauge theory
remain largely untested experimentally.
Careful studies of these couplings, for example in di-boson production
in $e^+e^-$ or hadronic collisions, may allow us not only to test
the  non-Abelian gauge structure of the SM~\cite{LAGRANGIAN},
but also to explore new physics at higher mass scales~\cite{WWVNEW}.

The reaction
$p\,p\hskip-7pt\hbox{$^{^{(\!-\!)}}$}  \rightarrow W^{\pm}\gamma$ is of
special interest due to the presence  of a zero in the amplitude of the
parton-level subprocess $q_1\bar q_2 \rightarrow W^\pm\gamma$
at~\cite{RAZ1}
\begin{eqnarray}
\cos\theta={Q_1+Q_2\over Q_1-Q_2}~,
\label{EQ:WGAMRAD}
\end{eqnarray}
where $\theta$ is the scattering angle of the $W$-boson with respect to the
quark ($q_1$) direction, and $Q_i$ ($i=1,2$) are the quark charges in
units of the proton electric charge $e$.
This zero is a consequence of the factorizability~\cite{RAZ2}  of
the amplitudes in gauge theories into one factor which contains the
gauge coupling dependence and another which  contains spin information.
Although the factorization holds for any four-particle Born-level
amplitude in which one or more of the four particles is a gauge-field
quantum, the amplitudes for most processes may not necessarily
develop a  kinematical zero in the physical region. The amplitude zero
in the $W^\pm \gamma$ process has been further shown to
correspond to the absence of dipole radiation by colliding particles
with the same charge-to-mass ratio~\cite{RAZ3}, a realization of
classical radiation interference. This phenomenon may make it possible
to measure the magnetic dipole moment and electric quadrupole moment of
the $W$-boson~\cite{VALENZUELA,UD}.

In this Letter we study in detail the helicity amplitudes of the
analogous process
\begin{eqnarray}
f_1(p_1) \> \bar f_2(p_2) \rightarrow W(p_{\rm w}) \> Z(p_{\rm z}) \>,
\label{EQ:REACT}
\end{eqnarray}
where $f_i$ represents a quark or lepton and $p_i$ denotes the particle
four-momentum. The three lowest order Feynman diagrams contributing
to the process $f_1 \bar f_2 \to W  Z$
are shown in Fig.~\ref{FIG:BORNGRAPHS}.
The helicity amplitudes for this process can be written as
\widetext
\begin{eqnarray}
{\cal M}(\lambda_{\rm w},\lambda_{\rm z}) =
F \, \epsilon_\mu^* (p_{\rm w},\lambda_{\rm w})
  \, \epsilon_\nu^* (p_{\rm z},\lambda_{\rm z}) \,
\overline{V}(p_2) \, T^{\mu\nu} \,
{1\over 2} \, (1-\gamma_5) \, U(p_1) \>,
\label{EQ:SMAMP}
\end{eqnarray}
\narrowtext
where $\lambda_{\rm w}$ ($\lambda_{\rm z}$) denotes the $W$ ($Z$) boson
polarization ($\lambda = \pm 1, 0$ for transverse and longitudinal
polarizations, respectively), $\epsilon$ denotes the weak-boson polarization
vector, and $U$ ($V$) represents the fermion (anti-fermion) spinor.
The factor $F$ contains coupling factors,
\widetext
\begin{eqnarray}
F = C \, {e^2 \over \sqrt{2} \sin \theta_{\rm w} } \>,
\end{eqnarray}
\narrowtext
where $\theta_{\rm w}$ is the weak mixing angle, and
$C$ is a color factor; for leptons $C=1$ and for
quarks $C = \delta_{i_1 i_2} \, V_{f_1 f_2}$.
Here $i_1$ ($i_2$) is the color index of the incoming
quark (antiquark) and $V_{f_1 f_2}$ is the quark mixing matrix element.
In the SM the tensor $T^{\mu \nu}$ is, in the limit of massless fermions,
\widetext
\begin{eqnarray}
T^{\mu \nu} &=& g_{-}^{f_1} \gamma^\mu
{(p\llap/_1 {-} p\llap/_{\rm z}) \over u} \gamma^\nu
+ g_{-}^{f_2} \gamma^\nu
{(p\llap/_1 {-} p\llap/_{\rm w}) \over t} \gamma^\mu
+ {Q_W\, \cot \theta_{\rm w} \over s - M_W^2}
\Bigl[ (p\llap/_{\rm z} {-} p\llap/_{\rm w}) \, g^{\mu\nu}
+ 2 p^{\nu}_{\rm w} \gamma^\mu
- 2 p^{\mu}_{\rm z} \gamma^\nu \Bigr] ,
\end{eqnarray}
\narrowtext
where the slash denotes a contraction with a Dirac gamma matrix,
$p\llap/ \equiv p_{\mu} \gamma^{\mu}$, $M_W$ is the $W$-boson mass,
$Q_W$ is the electric charge of the outgoing $W$-boson, and
$g^{f_i}_{-}$ are the couplings of the $Z$-boson to left-handed
fermions:
\widetext
\begin{eqnarray}
g_{-}^f = {1 \over \sin\theta_{\rm w} \cos\theta_{\rm w} } \>
(T_3^f - Q_f \sin^2\theta_{\rm w}) \>.
\label{EQ:COUPLING}
\end{eqnarray}
\narrowtext
In Eq.~(\ref{EQ:COUPLING}), $T_3^f = \pm {1 \over 2}$ represents the
third component of the
weak isospin and $Q_f$ is the electric charge of the fermion $f$.
The kinematic variables are defined by
\widetext
\begin{eqnarray}
s = (p_1+p_2)^2 \, ,  \,
t = (p_1-p_{\rm w})^2 = -{s \over 2}\,(\alpha - \beta \cos\theta) \, , \,
u = (p_1-p_{\rm z})^2 = -{s \over 2}\,(\alpha + \beta \cos\theta) \, ,
\end{eqnarray}
\narrowtext
where
\widetext
\begin{eqnarray}
\alpha = 1 - r_{\rm w} - r_{\rm z} \>, \qquad
\beta = ( \alpha^2 - 4 r_{\rm w} r_{\rm z} )^{1/2} \>, \qquad
r_{\rm v} = {M_V^2 \over s} \>,
\end{eqnarray}
\narrowtext
with $V=W$, $Z$, and $\theta$ is the center of mass scattering angle of the
$W$-boson with respect to the fermion ($f_1$) direction.
In the SM, the ${\rm SU_L(2)} \bigotimes {\rm U_Y(1)}$ gauge
structure relates the $Z$-boson to fermion couplings to the triple
gauge-boson $WWZ$ coupling by
\begin{eqnarray}
g^{f_1}_{-} - g^{f_2}_{-} = Q_W \, \cot\theta_{\rm w} \>.
\label{EQ:GG}
\end{eqnarray}

Using Eq.~(\ref{EQ:GG}) and the kinematic identity $s+t+u=M_W^2 +
M_Z^2$, the amplitude in Eq.~(\ref{EQ:SMAMP}) can be rewritten as
\widetext
\begin{eqnarray}
{\cal M}(\lambda_{\rm w},\lambda_{\rm z}) =
{F \over s}~ \overline{V}(p_2) \, \Biggl[
X \, \Bigl( A\llap/ - {t\over s-M_W^2} \, \Gamma\llap/ \Bigr)
+ Y \, \Gamma\llap/ \Biggr] \, (1-\gamma_5) \, U(p_1)  \>,
\label{EQ:FACTAMP}
\end{eqnarray}
\narrowtext
where $X$ and $Y$ are combinations of coupling factors
\begin{eqnarray}
X = {s \over 2} \, \left(  {g_{-}^{f_1} \over u}
                         + {g_{-}^{f_2} \over t} \right) \>, \qquad\qquad
Y = g_{-}^{f_1} \, {M_Z^2 \,s\, \over 2\,u\, (s - M_W^2) } \>,
\end{eqnarray}
and
\begin{eqnarray}
A\llap/ &=& 2\, p_1 {\cdot} \epsilon_{\rm w}^* \, \epsilon\llap/_{\rm z}^*
+\epsilon\llap/_{\rm z}^* \,\epsilon\llap/_{\rm w}^* \,p\llap/_{\rm w} \>,
\qquad \qquad
\Gamma\llap/ = 2 \Bigl[
  \epsilon_{\rm w}^* {\cdot} \epsilon_{\rm z}^* \,p\llap/_{\rm z}
+ p_{\rm w} {\cdot} \epsilon_{\rm z}^* \,\epsilon\llap/_{\rm w}^*
- p_{\rm z} {\cdot} \epsilon_{\rm w}^* \,\epsilon\llap/_{\rm z}^*
\Bigr] \>,
\end{eqnarray}
contain the spin dependent parts. $\Gamma\llap/$ acts as a projection
operator  on the
$J=1$ partial wave amplitudes. The form of the helicity amplitudes
given in Eq.~(\ref{EQ:FACTAMP}) is very convenient for studying the
factorization properties of the $WZ$ production amplitudes and for
comparing with the helicity amplitudes for $q_1\bar q_2 \rightarrow
W\gamma$. It is obvious that without the term  $Y$, which is
proportional to $M_Z^2$, the
helicity amplitudes would factorize. In this case all amplitudes would
simultaneously vanish for
$g_-^{f_1}/u + g_-^{f_2}/t =0$, in analogy to the $W\gamma$ case.

Further insight can be obtained from the explicit expressions for the helicity
amplitudes ${\cal M}(\lambda_{\rm w} , \lambda_{\rm z} )$.
Working in the parton center of mass frame, Eq.~(\ref{EQ:FACTAMP}) yields
\widetext
\begin{eqnarray}
\noalign{\vskip 5pt}
{\cal M}(\pm,\mp) &=& F\,  \sin\theta \,
\Bigl( \lambda_{\rm w} - \cos\theta \Bigr) X \>, \label{EQ:MPM} \\
\noalign{\vskip 8pt}
{\cal M}(\pm,\pm) &=& F\,  \sin\theta \, \left[ \left(
 \lambda_{\rm w} (r_{\rm z} - r_{\rm w})  -\beta + \cos\theta
+ \beta \, { \alpha - \beta \cos\theta \over 1 - r_{\rm w} }   \right) X
+ 2 \beta Y \right] \>, \label{EQ:MPP} \\
\noalign{\vskip 8pt}
{\cal M}(0,0) &=& F\,
{\sin\theta \over \sqrt{2 r_{\rm w} 2 r_{\rm z}}} \,
\left[ \left( -\beta\rho
+ \beta_{\rm w} \beta_{\rm z} \cos\theta
+ \beta\rho  \, { \alpha - \beta \cos\theta \over 1 - r_{\rm w} } \right) X
+ 2 \beta \rho Y \right] \>, \\
\noalign{\vskip 8pt}
{\cal M}(\pm,0) &=& F\,
{(1 - \lambda_{\rm w} \cos\theta) \over \sqrt{2 r_{\rm z}} }\,
\left[ \left( 2  \lambda_{\rm w} r_{\rm z}  -\beta
+ \beta_{\rm z} \cos\theta
+ \beta \, { \alpha - \beta \cos\theta \over 1 - r_{\rm w} } \right) X
+  2 \beta Y \right] \>, \\
\noalign{\vskip 8pt}
{\cal M}(0,\pm) &=& F\,
{(1 + \lambda_{\rm z} \cos\theta) \over \sqrt{2 r_{\rm w}} }\,
\left[ \left(  - 2  \lambda_{\rm z} r_{\rm w}   -\beta
+ \beta_{\rm w} \cos\theta
+ \beta \, { \alpha - \beta \cos\theta \over 1 - r_{\rm w} } \right) X
+  2 \beta Y \right] \>,
\label{EQ:MZEROPM}
\end{eqnarray}
\narrowtext
where $\beta_{\rm w}$, $\beta_{\rm z}$, and $\rho$
are combinations of the kinematic invariants,
\widetext
\begin{eqnarray}
\beta_{\rm w}  = 1 + r_{\rm w} - r_{\rm z} \>, \qquad
\beta_{\rm z}  = 1 - r_{\rm w} + r_{\rm z} \>,  \qquad
\rho           = 1 + r_{\rm w} + r_{\rm z} \>.
\end{eqnarray}
\narrowtext

The amplitudes given in Eqs.~(\ref{EQ:MPM})~--~(\ref{EQ:MZEROPM})
contain the main results of our paper. They exhibit several interesting
features. ${\cal M}(\pm,\mp)$ receive contributions only from $J\ge 2$
partial waves, {\it i.e.}, only from the $u$ and $t$-channel
fermion-exchange diagrams (see Fig.~\ref{FIG:BORNGRAPHS}a and~b).
The $(\pm,\mp)$ amplitudes therefore do not depend on
$Y$ and thus factorize. They vanish for
\widetext
\begin{eqnarray}
{g^{f_1}_{-} \over u} + {g^{f_2}_{-} \over t} = 0 \>, \hskip 0.7cm {\rm
or}  \hskip 0.7cm \cos\theta_{0} =  {\alpha \over \beta} \,
\left( {g^{f_1}_{-} + g^{f_2}_{-} \over g^{f_1}_{-} - g^{f_2}_{-} } \right)
\>.
\end{eqnarray}
\narrowtext
The existence of the zero in ${\cal M}(\pm,\mp)$ at $\cos\theta_0$ is
a direct consequence of the contributing Feynman diagrams and the
left-handed coupling of the $W$-boson to fermions.
Unlike the $W^\pm \gamma$ case with its massless photon kinematics, the
zero has an energy dependence
through  $\alpha$ and $\beta $ which is, however, rather weak for
energies sufficiently above the $WZ$ mass threshold. More explicitly, for
$s \gg M_V^2$, the zero is located at
\[ \cos\theta_0 \simeq \left\{
\begin{array}{ll}
\phantom{+}{1\over 3}\tan^2\theta_{\rm w} \simeq 0.1
& \mbox{for $d \bar u \rightarrow W^{-} Z\>,$} \\
 - \tan^2\theta_{\rm w}  \simeq -0.3
& \mbox{for $e^{-} \bar \nu_e \rightarrow W^{-} Z\>,$}  \\
-{1\over 3}\tan^2\theta_{\rm w}  \simeq  -0.1
& \mbox{for $u \bar d \rightarrow W^{+} Z\>,$} \\
\phantom{+} \tan^2\theta_{\rm w}  \simeq  0.3
& \mbox{for $\nu_e e^{+} \rightarrow W^{+} Z\>.$}
\end{array}
\right. \]

At high energies, $s \gg M_V^2$, strong cancelations occur
between the various terms in the square brackets in
Eqs.~(\ref{EQ:MPP})~--~(\ref{EQ:MZEROPM}) and only
the $(\pm,\mp)$ and $(0,0)$ amplitudes remain non-zero:
\widetext
\begin{eqnarray}
{\cal M}(\pm,\mp) &\longrightarrow&  {F \over \sin\theta }\,
(\lambda_{\rm w} - \cos\theta)\,
\Bigl[  (g^{f_1}_{-} - g^{f_2}_{-} ) \cos\theta
      - (g^{f_1}_{-} + g^{f_2}_{-} ) \Bigr] \>, \\
\noalign{\vskip 10pt}
{\cal M}(0,0) &\longrightarrow&  {F\over 2} \, \sin\theta \,
{M_Z \over M_W} \, (g^{f_2}_{-} -  g^{f_1}_{-}) \>.
\end{eqnarray}
\narrowtext
Note that the amplitudes do not fully factorize in the high energy
limit due to the survival of ${\cal M}(0,0)$. Because the $Y$ term in
Eq.~(\ref{EQ:FACTAMP}) behaves like
$(M_Z^2/s)\, \epsilon^*_{\rm w} \epsilon^*_{\rm z}$ at high energies,
one might naively expect that the amplitudes completely factorize for $s\gg
M_V^2$ as in the $W\gamma$ case. However, for longitudinally polarized
vector bosons $\epsilon^*_{\rm v} \sim \sqrt{s}/M_V^{}$ and thus the $(0,0)$
amplitude remains finite.

The combined effect of the zero in ${\cal M}(\pm,\mp)$ and the gauge
cancelations at high energies in the remaining helicity amplitudes
results in an approximate zero for the $f_1\bar f_2\rightarrow W^\pm Z$
differential cross section at $\cos\theta\approx\cos\theta_0$. This is
illustrated in Fig.~\ref{FIG:AZERO} where we show the
differential cross sections for
$e^-\bar\nu_e \rightarrow W^-Z$ and $d\bar u\rightarrow W^-Z$,
\begin{eqnarray}
{d\sigma(\lambda_{\rm w},\lambda_{\rm z})\over d\cos\theta} =
{\beta \over 32 \pi s} \
\overline{ |{\cal M}(\lambda_{\rm w},\lambda_{\rm z})|^2 } \>,
\label{EQ:SIGMA}
\end{eqnarray}
(the over-line denotes the fermion-spin and color averaged squared
matrix element)
for $(\lambda_{\rm w},\lambda_{\rm z})=(\pm,\mp)$ and $(0,0)$,
as well as the unpolarized cross
section, which is obtained by summing over all $W$- and $Z$-boson helicity
combinations (solid line).
Although the matrix elements are calculated at $\sqrt s = 2$~TeV,
the results differ little from those obtained in the high energy limit.
For both reactions, the total differential cross section displays a
pronounced minimum at
the location of the zero in ${\cal M}(\pm,\mp)$. Due to the
$1/\sin\theta$ behaviour of ${\cal M}(\pm,\mp)$, the $(+,-)$ and $(-,+)$
amplitudes dominate outside of the region of the zero. In order to
demonstrate the influence of the zero in ${\cal M}(\pm,\mp)$ on the
total angular differential cross section, we have included the
$\cos\theta$ distribution for $e^+e^-\rightarrow ZZ$ in
Fig.~\ref{FIG:AZERO}a, normalized to the
$e^-\bar\nu_e \rightarrow W^-Z$ cross section at $\cos\theta=0.9$
(long dashed line). The zero in the $(\pm,\mp)$
amplitudes causes the minimum in the $WZ$ case to be much more
pronounced than the minimum in $e^+e^-\rightarrow ZZ$.

Figure~\ref{FIG:DSIGMADCOS} illustrates the energy dependence of the
differential cross section. At $\sqrt{s}=0.2$~TeV, {\it i.e.} close to
the threshold, contributions from the $(\pm,\pm)$, $(\pm,0)$,
and $(0,\pm)$ amplitudes are all important. These amplitudes do not
factorize in terms of $X$ and therefore tend to eliminate the
approximate zero (dotted lines). Above threshold, these contributions
rapidly diminish, as exemplified
by the curves for $\sqrt{s}=0.5$~TeV (dashed lines) and $\sqrt{s}=2$~TeV
(solid lines). Note that the location of the minimum varies only
slightly with energy.

The results obtained so far demonstrate that the SM $WZ$ cross section
exhibits an approximate kinematical zero in the physical region in the
high energy limit.  This behavior is due to
the combined effect of an exact zero in ${\cal M}(\pm,\mp)$ and strong
cancelations in the remaining amplitudes.
Non-standard $WWZ$ couplings spoil these cancelations and
eliminate the approximate zero. To illustrate this,
we consider the general $C$ and $P$ conserving effective
Lagrangian~\cite{LAGRANGIAN}
\begin{eqnarray}
{\cal L}_{WWZ} &=& -i \, e \, {\rm cot} \theta_{\rm w}
\Biggl[ g_1\, \bigl( W_{\mu\nu}^{\dagger} W^{\mu} Z^{\nu}
                  -W_{\mu}^{\dagger} Z_{\nu} W^{\mu\nu} \bigr)
+ \kappa \, W_{\mu}^{\dagger} W_{\nu} Z^{\mu\nu}
+ {\lambda \over M_W^2}\,W_{\lambda \mu}^{\dagger} W^{\mu}_{\nu}
Z^{\nu\lambda}
\Biggr] \>,
\label{EQ:LAGRANGE}
\end{eqnarray}
\narrowtext
where $W^{\mu}$ and $Z^{\mu}$ are the $W^-$ and $Z$ fields, respectively, and
$V_{\mu\nu} = \partial_{\mu}V_{\nu} - \partial_{\nu}V_{\mu}$, $V=W,\,Z$.
In the SM at tree level, $g_1 = 1$, $\kappa = 1$, and $\lambda = 0$.
The contributions to the $WZ$ production amplitudes
from the anomalous couplings $\Delta g_1 = g_1 - 1$,
$\Delta\kappa = \kappa -1$, and $\lambda$ are
\widetext
\begin{eqnarray}
\Delta{\cal M}(\pm,\mp) &=& 0\>,
\label{EQ:PLUSMIN} \\
\noalign{\vskip 8pt}
\Delta{\cal M}(\pm,\pm) &=& {F\over 2} \>
{Q_W \, \cot\theta_{\rm w} \over 1 - r_{\rm w} } \> \beta \sin\theta \,
\Biggl[ \Delta g_1 + \Delta\kappa + {\lambda \over r_{\rm w} } \Biggr] \>,\\
\noalign{\vskip 8pt}
\Delta{\cal M}(0,0) &=& {F\over 2} \>
{Q_W \, \cot\theta_{\rm w} \over 1 - r_{\rm w} } \>
{\beta \sin\theta \over \sqrt{2 r_{\rm w} 2 r_{\rm z}} } \, 2 \,
\Bigl[ \Delta g_1 (1 + r_{\rm w}) + \Delta\kappa\, r_{\rm z} \Bigr] \>,
\label{EQ:ZEROZERO} \\
\noalign{\vskip 8pt}
\Delta{\cal M}(\pm,0) &=& {F\over 2} \>
{Q_W \, \cot\theta_{\rm w} \over 1 - r_{\rm w} } \>
{\beta (1- \lambda_{\rm w}\cos\theta ) \over \sqrt{2 r_{\rm z}} } \,
\Biggl[ 2\, \Delta g_1 + \lambda \, {r_{\rm z}\over r_{\rm w}} \Biggr] \>,\\
\noalign{\vskip 8pt}
\Delta{\cal M}(0,\pm) &=& {F\over 2} \>
{Q_W \, \cot\theta_{\rm w} \over 1 - r_{\rm w} } \>
{\beta (1 + \lambda_{\rm z}\cos\theta ) \over \sqrt{2 r_{\rm w}} } \,
\Bigl[ \Delta g_1 + \Delta\kappa + \lambda \Bigr] \>.
\label{EQ:NSMHELAMP}
\end{eqnarray}
\narrowtext
The amplitudes given in Eqs.~(\ref{EQ:PLUSMIN}) --~(\ref{EQ:NSMHELAMP})
agree with those of Refs.~\cite{BZ} apart from terms proportional to
$(M_Z^2-M_W^2)/s$ in $\Delta{\cal M}(0,0)$. The expression given in
Eq.~(\ref{EQ:ZEROZERO}) corrects a minor error in Ref.~\cite{BZ}.

Due to angular momentum conservation, the $(\pm,\mp)$ amplitudes which
dominate in the SM do not receive any contributions from the
anomalous couplings. The amplitude zero
in these channels thus remains exact. All other helicity amplitudes are
modified in the presence of non-standard $WWZ$ couplings. At
high energies the anomalous contributions grow proportional to
$\sqrt{s}$ ($s$) for $\Delta\kappa$ ($\Delta g_1$ and $\lambda$)
and thus could dominate the cross section. $\lambda$, $\Delta g_1$, and
$\Delta\kappa$ contribute predominantly to the
$(\pm,\pm)$, $(0,0)$, and $(0,\pm)$ amplitudes, respectively~\cite{ZW}.
The influence of non-standard $WWZ$ couplings on the differential cross
section is illustrated in Fig.~\ref{FIG:ANOMALOUS}, where we compare
$d\sigma/d\cos\theta$ for $d\bar u\rightarrow W^-Z$ at
$\sqrt{s}=500$~GeV for $\lambda=0.1$, $\Delta g_1=0.2$, and
$\Delta\kappa=0.5$ (solid lines) with the SM result (dashed lines).
Besides the total cross section, we also display the cross section for
the helicity state which
produces the largest non-standard contribution for the given anomalous
coupling.
In each case displayed, the approximate zero present in the SM is
completely eliminated by the anomalous coupling.

We finally comment briefly on observable signals of the approximate
zero in $WZ$ production in the SM. $e\nu_e$ or $\mu\nu_\mu$ collisions
above the $WZ$ threshold would in principle provide a clean environment,
and the location of the zero at $\cos\theta_0\approx \pm 0.3$ is ideal
for experimental studies of the $W^\pm Z$ final state, unlike the case for
$e^-\bar \nu_e \rightarrow W^-\gamma$ where the zero is
located at the kinematical boundary, $\cos\theta=1$~\cite{RAZ1}.
In the foreseeable future $WZ$ production can be studied
most easily in hadronic collisions. Since the
high energy limit of the helicity amplitudes is approached very rapidly
above threshold, the approximate amplitude zero in the SM results in a dip
in the distribution of the $Z$-boson rapidity in the parton center of
mass frame in $p\, p\hskip-7pt\hbox{$^{^{(\!-\!)}}$}\rightarrow WZ$, in
complete analogy to the dip in the photon center of
mass rapidity distribution which signals the radiation zero in $W\gamma$
production~\cite{UD}. Alternatively, one can study $W$-$Z$ rapidity
correlations where a similar dip occurs~\cite{FRIX}.

In summary, we have demonstrated that the amplitude for
$f_1 \bar f_2 \rightarrow  W^\pm Z $ at Born-level exhibits an
approximate zero in the SM in the physical region at high energies.
The approximate zero
is the combined result of an exact zero in ${\cal M}(\pm,\mp)$ and
strong gauge cancelations in the remaining helicity amplitudes. The zero
in the $(\pm,\mp)$ amplitudes is a direct consequence of the
contributing Feynman diagrams and the left-handed coupling of the
$W$-boson to fermions. Anomalous
$WWZ$ couplings spoil the gauge cancelations and eliminate the effect. The
approximate amplitude zero leads to a pronounced dip in the $Z$-boson
center of mass rapidity distribution in $p\,
p\hskip-7pt\hbox{$^{^{(\!-\!)}}$}\rightarrow WZ$, which
in principle can be observed experimentally.


%
\acknowledgements

We would like to thank T.~Fuess, C.~Wendt, and D.~Zeppenfeld for
stimulating discussions.
This work has been supported in part by Department of Energy grants
\#DE-FG03-91ER40674 and \#DE-FG05-87ER40319
and by Texas National Research Laboratory grant \#RGFY93-330.
T.H. is also supported in part by a UC-Davis Faculty Research Grant.
%
%

%
%
%
\begin{figure}
\caption{Feynman diagrams contributing to the Born-level subprocess
$f_1 \bar f_2 \rightarrow W Z$.}
\label{FIG:BORNGRAPHS}
\end{figure}
\begin{figure}
\caption{Differential cross section $d\sigma(\lambda_{\rm w},
\lambda_{\rm z})/d\cos\theta$
versus the  $W^-$ scattering angle $\theta$ in the center of mass frame
for the Born-level processes  (a) $e^- \bar
\nu_e \rightarrow W^-Z $ and (b) $d \bar u \rightarrow W^-Z $.
The dashed, dotted, and dash-dotted curves are for
$(\lambda_{\rm w},\lambda_{\rm z})=(0,0)$, $(+,-)$, and $(-,+)$,
respectively. The solid line represents the total (unpolarized) cross
section. For comparison, the long dashed curve in (a) shows the
$e^+e^-\rightarrow ZZ$ cross section, normalized to the
$e^-\bar\nu_e \rightarrow W^-Z$ cross section at $\cos\theta=0.9$.}
\label{FIG:AZERO}
\end{figure}
\begin{figure}
\caption{Differential cross section $d\sigma/d\cos\theta$
versus   the  $W^-$ scattering angle $\theta$ in the center of mass frame
for the Born-level processes  (a) $e^- \bar \nu_e \rightarrow W^-Z $
and (b) $d \bar u \rightarrow W^-Z $. The dotted, dashed, and solid
curves are for  $\protect{\sqrt{s}=0.2}$, 0.5, and 2~TeV, respectively.}
\label{FIG:DSIGMADCOS}
\end{figure}
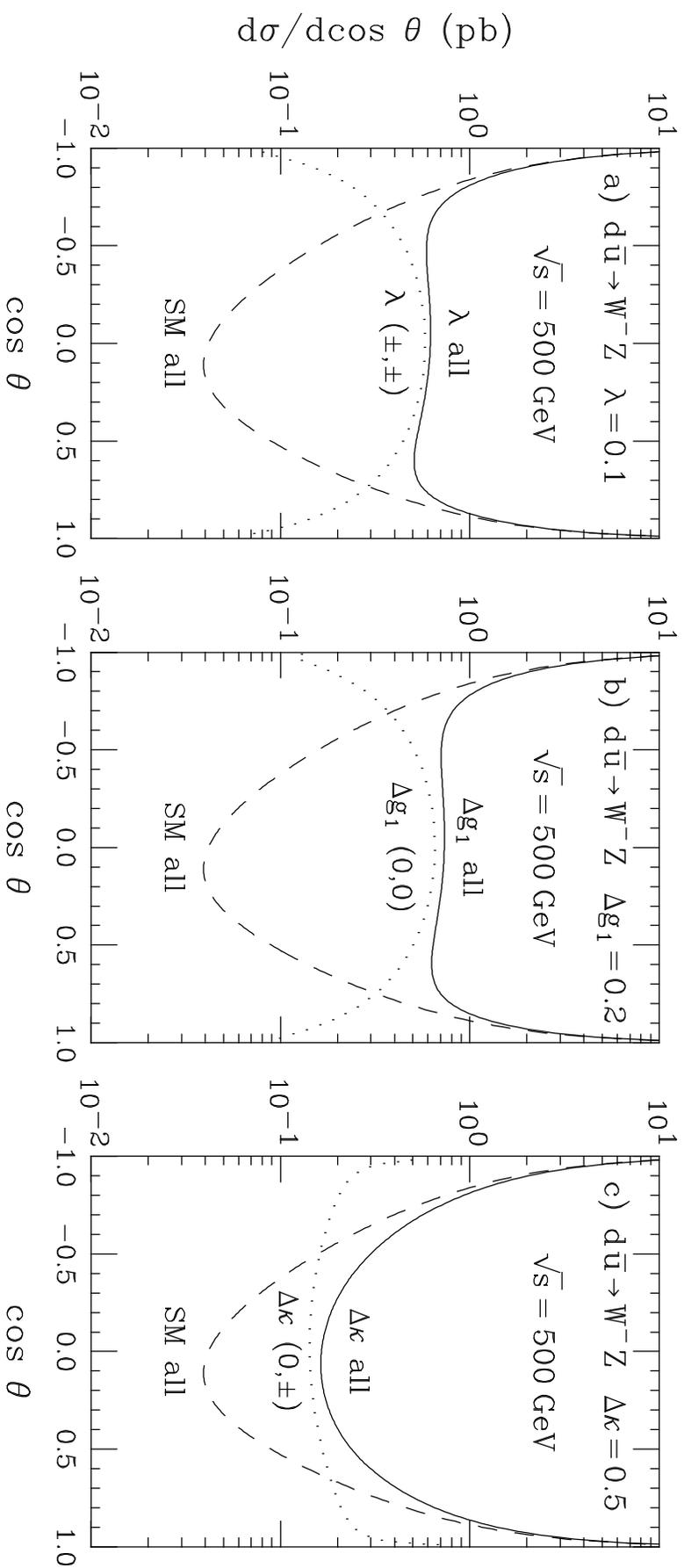
\begin{figure}
\caption{Differential cross sections for the subprocess
$d \bar u \rightarrow W^-Z$ at $\protect{\sqrt{s}=500}$~GeV with
anomalous couplings as defined in Eq.~(\protect{\ref{EQ:LAGRANGE}}).
Parts a), b), and c) are for $\lambda = 0.1$,
$\Delta g_1 = 0.2$, and $\Delta \kappa = 0.5$, respectively. The solid
(dashed) lines give the total differential cross section with (without)
anomalous $WWZ$ couplings. The dotted curves show the cross section for
the helicity state which produces the largest non-standard contribution
for the given anomalous coupling.}
\label{FIG:ANOMALOUS}
\end{figure}
%
%
\end{document}